\documentclass{aa}  
\usepackage{graphicx}
\usepackage{natbib}
\bibpunct{(}{)}{;}{a}{}{,}

\begin{document}

\authorrunning{I.~Ermolli et al.}
\titlerunning{Potential value Ca~II~K spectroheliogram time-series}
   \title{On the potential value of Ca~II~K spectroheliogram time-series\\ for solar activity and irradiance studies}

   \author{I.~Ermolli\inst{1}, S.~K.~Solanki\inst{2}, A.~G.~Tlatov\inst{3}, N.~A.~Krivova\inst{2},  R.~K.~Ulrich\inst{4} \and J.~Singh\inst{5}}
      
   \offprints{Ilaria Ermolli \email{ermolli@oaroma.inaf.it}}

   \institute{ INAF - Osservatorio astronomico di Roma, Via Frascati 33, 00040 Monte Porzio Catone, Italy
\and  
Max-Planck-Institut f\"ur Sonnensystemforshung, Max-Planck-Strasse 2, 37191 Katlenburg-Lindau, Germany 
\and
Kislovodsk Solar Station, Pulkovo Observatory, Pulkovskoe Ch. 65-1, 196140 Saint Petersburg, Russia
\and
Division of Astronomy and Astrophysics, University of California, 8371 Mathematical Science Building, Los Angeles, CA 90095-1562, USA
\and
Indian Institute of Astrophysics, 2nd Block, Koramangala, 560034 Bangalore, India
}
   \date{}

   \abstract
 {Various observatories around the globe started regular full-disk imaging of the solar  atmosphere in the Ca~II~K line 
 since the early decades of the 
 20th century. 
The archives made by these observations have the potential of providing far more detailed information on solar magnetism  
than just 
the sunspot number and area records to which most studies of solar activity and irradiance changes are restricted. 
}
{We evaluate the image contents 
of three  Ca~II~K spectroheliogram time-series, specifically those obtained by the digitization of the 
Arcetri, Kodaikanal, and Mt Wilson photographic
archives, in order to estimate their value for studies focusing on time-scales longer than the solar cycle. 
}   
{We describe the main problems 
afflicting these data  and analyze their quality  by expressing the image contents through several 
quantities. We compare the results obtained with those for similar present-day  observations  
taken with the Meudon spectroheliograph and with the Rome-PSPT.} 
{We show that 
historic data suffer from stronger geometrical distortions and photometric uncertainties
than similar present-day observations. The latter uncertainties mostly originate from the photographic calibration 
of the original data and from stray-light effects.    
We also show that the image contents  of the three analyzed series vary in time. 
These variations are probably due to instrument changes and  aging of 
the spectrographs used, as well as changes of the observing programs. 
Our results imply 
that the main challenge for the analysis of historic data is their accurate photometric calibration. This problem must be solved before they can 
provide 
reliable information about solar magnetism and activity
over the last century. Moreover, inter-calibration of results obtained from independent time-series is required to reliably trace changes
of solar properties with time from the analysis of such data.}
{}


\maketitle

\section{Introduction}
A wide variety of solar research,  ranging from the investigation of global solar activity
and variability to the study of large scale patterns of proper motions, 
 is based upon the analysis of regular full-disk observations of the Sun.  Only during the last solar cycle have such 
 observations been carried out by space-based telescopes  and by the new generation of 
ground-based  instruments, e.g. by SoHO/MDI \citep{scherrer1995}, PSPT \citep{coulter1994, ermolli1998}, 
CFDT2-SFO \citep{chapman2004}, and SOLIS \citep{kellerharvey2003}. These observations  
are thus of limited usefulness for  focussing on time-scales longer than the activity cycle. For such studies
regular full-disk observations of 
the solar atmosphere starting at the beginning  of the 20th century at several observatories are of particular interest 
 \citep[for a list of synoptic programs carried out before 1950 see][]{mouradian2007}.  These historic observations
were made in white light and in various spectral bands, often in the Ca~II~K and H$\alpha$ resonance lines, 
mostly using spectroheliographs. 

Among the historic series, those including Ca~II~K observations have the largest potential 
of providing  information about  
solar magnetism. In fact, Ca~II~K emission can be used as a good
proxy of the line-of-sight magnetic flux density 
\citep[][]{skumanich1975,schrijver_etal1989,ortiz_rast2005}. 
Note that in standard notation {\it $K_3$, $K_{2V,2R}$}, and {\it $K_{1{\rm V},1{\rm R}}$} mark the core, 
the reversal, and the secondary 
minimum of the doubly-reversed profile of the Ca~II~K line, in the violet ({\rm V}) and the red ({\rm R}) wings of the line, respectively. 
All these 
line features occur within a spectral range less than 1~\AA 
~wide. 

In principle,  historic  Ca~II~K observations constitute an extremely 
valuable resource for many research topics. However, to date 
analysis of the Ca~II~K spectroheliogram time-series was restricted for two reasons: 1) lack of data in digital format; 2) shortcomings and defects 
that beset
 historic data. The first restriction should be overcome soon by the results of new projects devoted to the digitization of some of the major photographic 
archives. For instance, Arcetri, Kodaikanal and Mt Wilson Ca~II~K historic observations have already been  digitized 
\citep{ulrich2004,makarov2004,marchei2006}, and other similar series 
are now being processed as well. 
Defects in and decay of spectroheliogram photographic plates, missing photographic 
calibration and undocumented changes of the used instrumentation, however, lead to various artifacts in and problems with the historic data 
 \citep{zharkova_etal2003,fuller2005,ermolli2007}, which are avoided in the full-disk 
images taken by the most recent synoptic observing programs.
 
Here we intercompare and discuss three time-series of images obtained by the digitization of Ca~II~K historic spectroheliograms.
 We describe the main problems 
affecting these data (Sect.~2) and analyze their quality (Sect.~3). In particular, we measure the image contents of these data 
and compare the results with those obtained for similar present-day observations. 
The objective of this study is twofold. First of all, it should help to estimate the value of Ca~II~K historic observations for  
studies of solar activity, magnetism and irradiance on time-scales longer than the activity cycle (Sects. 4, 5). Secondly, it should raise 
our understanding of the characteristics that an appropriate image processing technique must have
for uniform and accurate application to such data. We see this study as the first step towards a systematic exploitation of this valuable resource.

\section{Data, definitions and pre-processing}
\subsection{Data description}
The current analysis concentrates mainly on images obtained from the digitization of Ca~II~K  spectroheliograms stored in the photographic
archives of the Arcetri, Kodaikanal 
and Mt Wilson Observatories (henceforth referred to as historic data). In 
addition, we have also analyzed 
samples of Ca~II~K images obtained by  two current synoptic programs, namely those carried out with the 
Meudon spectroheliograph and with the Rome-PSPT telescope (henceforth referred to as modern data). 

{\it Historic data -} The first set of spectroheliograms analyzed here was recorded at the G.~B.~Donati tower of the INAF Arcetri Astrophysical Observatory in Florence from
1931 to 1974 ({\it Ar}, hereafter).
The  
spectroheliograph used for these observations \citep{gr1950,ga2004} had a grating of 600~lines/mm and a ruled area of 
100~$\times$~110~mm, with a  
dispersion of 0.33~mm/\AA ~at 3934~\AA. The size of the solar disk on most  plates is $\approx$~6.5~cm; the image scale is thus 
about 0.033~mm/$\arcsec$.
The spectral window for these observations was 0.3~\AA 
~centered in the line core.
The instrumentation used to acquire these observations is no longer available.

It is worth noting that several instrumental changes occurred during the
over forty years that the spectroheliograph was utilized.
These include the use of additional lenses and changes of the slits positions, which improved the image definition and  monochromaticity, and
 decreased the 
stray-light level. The problem of discontinuities marked by instrumental changes is common to most, if not all, the existing 
long time-series of synoptic observations. The effects of instrumental changes upon the quality of the data analyzed in this study are presented 
in Sect.~3.

The digitization of the Arcetri archive was performed by the CVS project 
at the Rome Astronomical Observatory \citep{centrone2005,giorgi2005,marchei2006}. 
The work was carried out with a commercial scanner, 
used with the setting 1200~$\times$~1200~dpi and 16~bit gray-scale significant data.
From these data, which are saved in the TIFF format, 2040~$\times$~2720
 16~bit pixel FITS images for each solar observation were produced.
Information about the plate acquisition noted in the 
observation log-books was included in the FITS headers. 
The diameter of the solar disk in these images is about 1550~pixels; the pixel scale due to the digitization is thus $\approx$~1.2$\arcsec$/pixel.
The digital archive of Ca~II~K Arcetri observations  contains
5976 spectroheliograms obtained on 5042 observing days. The data analyzed here are the reduced-size (2~$\times$~binned) FITS format 
images available through the CVS 
archive\footnote{http://cvs3.mporzio.astro.it/$\sim$cvs/cvs/arcetri.html}. We analyzed 4052 images obtained on 3927 observing days.

The second set of spectroheliograms is the one recorded at the Kodaikanal Observatory of the Indian Institute of Astrophysics 
in Bangalore from 1907 to 1999 ({\it Ko}, hereafter). 
 The spectrograph  used for these observations 
\citep{evershed1911,bappu1967} is a two prism instrument, with a dispersion of 0.14~mm/\AA ~near 3930~\AA.
The 70~$\mu m$  exit slit of the 
instrument corresponds to a 0.5~\AA ~bandpass, which includes K$_{232}$.
The size of the solar disk on most  plates is $\approx$~6~cm; the image scale is thus 
about 0.031~mm/$\arcsec$. The instrument used to acquire these observations is still available.

The digitization of the Kodaikanal Ca~II~K archive \citep{makarov2004} was performed using a commercial scanner, used 
with the setting 1200~$\times$~1200~dpi and 8~bit 
gray-scale significant data. 
The original observations were stored as
$\approx$~1800~$\times$~1800 8~bit gray-scale JPEG images.
The solar disk in the digitized images has a diameter of about 1420~pixels; the pixel scale due to the digitization is thus 1.3$\arcsec$/pixel.
The digital archive of  Kodaikanal Ca~II~K observations contains 26640 spectroheliograms obtained  
on 26620 observing days. We analyzed here 19522 images  obtained on 19172 observing days.

The third set of spectroheliograms we analyzed is the one taken at the Mt Wilson Observatory from 1915 to 1985 ({\it MW}, hereafter). 
The  
spectroheliograph used at the beginning of these observations \citep{ellerman1919} had a grating with  590~lines/mm 
and a ruled area of 61.5~$\times$~79~mm. After the first decade of observation this grating was changed, and the ruled area was increased 
to 103~$\times$~107~mm. 
The 80~$\mu m$ exit slit admitted the passage of a band of about 0.2~\AA
~centered on the Ca~II~K line core.
The size of the solar disk on most  plates is 
 $\approx$~5~cm; the image scale is thus 
about 0.026~mm/$\arcsec$.

The Mt Wilson  observations were digitized in the framework of a project  carried out at UCLA Division of 
Astronomy \citep{ulrich2004,lefebvre2005}. The work was performed with 
a commercial scanner, used with the setting 1200~$\times$~1200~dpi and 16~bit gray-scale significant data. From these data, 
$\approx$~2600~$\times$~2600~pixels 16~bit TIFF images were 
singled out, in 
which intensity values are stored
as positive integers. The solar disk in these images has a diameter of about 2000 pixels, the pixel scale due to the digitization is 
thus about 1$\arcsec$/pixel. 
Information about log-book notes, plate dimensions, and contents is available on the project 
web-page\footnote{http://www.astro.ucla.edu/$\sim$ulrich/MW$_{-}$SPADP/}.

The  bulk of the {\it MW} images we analyzed are the reduced-size (800~$\times$~800~pixels) science quality FITS files posted for 
distribution on the UCLA project web-page. 
The file headers of these images contain information
about the acquisition of the original plate and its digitization, as well as measurements of the disk horizontal and vertical radii. 
These images were also processed in order to remove small-scale plate inhomogeneities, as described in 
the project web-page. We analyzed 34166 images obtained on 20684 observing days.

In addition, 
 we have also analyzed two other samples of {\it MW} images. The first sample contains 237 full-size images, which 
have thus $\approx$~2600~$\times$~2600~pixels each,  taken in the month of July 
from 1920 to 1930.  These images are 16-bit FITS files. 
The file headers of these images contain information about 
the radius and disk center of the solar observations.  The analyzed images were also processed by the UCLA scientists to remove small-scale 
inhomogeneities.   The analysis of this sample  aims at evaluating the sensitivity of  
the results to the reduction of the image size. The other sample analyzed contains 
713 reduced-size images (800~$\times$~800~pixels) which were photographically calibrated by the UCLA project scientists. These 
images were obtained by applying a calibration method based on the one presented by \citet{devaucouleurs1968} with some modifications. This method 
makes use of calibrated exposures made on the plates outside the solar disk for the data acquired since late 1961  
(details available on the project web-page). This sample contains  {\it MW} observations taken in 1967 and 1975.

{\it Modern data -} We also analyzed a sample of Ca~II~K spectroheliograms 
obtained at the Observatory of Paris-Meudon. These data ({\it Me}, hereafter) consist of 1044 full-disk images taken 
with the updated version of the instrument installed at the beginnig of the 20th century \citep{deslandres1891,deslandres1913}.  
This is a prism-spectroheliograph  with a dispersion of 0.48~mm/\AA ~at $K_{1{\rm V}}$ and $K_3$. 
 The size of the solar disk image is $\approx$~6.5~cm; the image scale is thus 
about 0.033~mm/$\arcsec$.
Images are acquired with a 1300~$\times$~1300 14~bit pixel CCD camera, with a pixel scale of $\approx$~1.5$\arcsec$/pixel. The
 spectral pass-band of these images is
0.15~\AA 
~centered on the $K_3$ and the $K_{1{\rm V}}$ line features (J.~M. Malherbe, private communication). 
We analyzed 903  $K_3$ and 141  $K_{1{\rm V}}$ images, respectively. These images were obtained on 87  observing days
in the month of July from 2004 to 2006. The data analyzed are the FITS format images available through the BASS2000 
archive\footnote{http://bass2000.obspm.fr}.

Finally, we also analyzed a sample of Ca~II~K full-disk images
obtained  with the PSPT telescope at the Rome Observatory \citep{ermolli2003}. These data ({\it PSPT}, hereafter) were recorded 
with a telescope designed  to provide high-precision photometric observations in several spectral bands \citep{coulter1994,ermolli1998}. 
Briefly, images are acquired with a 2048~$\times$~2048   12~bit pixel CCD camera, 2~$\times$~binned, with a final pixel scale 
of $\approx$~2$\arcsec$/pixel, using an interference filter centered
on the Ca~II~K line (393.3~nm, FWHM 0.25~nm). In particular, we analyzed 4448 images obtained on 2838 observing days from January 1998 up to December 2006.
The sample of data analyzed is composed of single frame images acquired with short exposure times, as well as 
images obtained by co-adding 25  frames acquired with short exposure times.  In general, the addition of frames,
which is aimed at a reduction of the photometric  measurement noise, also reduces the spatial resolution on images. The data analyzed are the FITS 
format images available through the Rome-PSPT archive\footnote{http://www.mporzio.astro.it/solare}.

 

\subsection{Definitions}
Values for each pixel (pixel value, $PV$ hereafter) of the {\it Ar, Ko}, and {\it MW} images  were provided by the scanning devices 
and measure the flux 
of the scanner beam transmitted through the photographic plate. Note that  the {\it MW} data available 
on the project web-page were reversed, in order to show bright solar features with higher pixel values. 
However, the pixel values of the {\it MW} images analyzed in this study describe the numbers 
provided by the scanner device. Given a proper scanner  calibration, $PV$ for each image pixel is a measure of the blackening degree 
of the photographic plate at the position corresponding to that pixel.
The  blackening is linked to the flux of solar radiation incident during the plate
exposure  by a relation which depends on many plate characteristics \citep{dainty1974}.  
The knowledge of this relation, which is called characteristic curve, allows a  calibration of the photographic blackening  into 
intensity. 
Note that some of the  image properties considered in this study do not depend on the photographic calibration 
of the original data. In particular, information on both geometrical distortions and spatial scale,  can be
obtained by analyzing photographically un-calibrated historic data. 
However, the evaluation of the photometric properties of the images, such as
stray-light level and contrast, requires photographic calibration of the data. This holds also for
the direct comparison between results obtained from historic and modern data.
 
Since only a small fraction of the historic images analyzed contains calibration exposures, we performed the photographic calibration of all these 
data  using a method independent of calibration exposures. 
Namely, we have converted the image pixel value  $PV$ 
of the {\it Ar, Ko}, and {\it MW} images 
into relative calibrated intensity values according to the 
formula \citep{burkanian}: $I_i=(V-B)/(T_i-B)$, where $I_i$ is the calibrated intensity (in arbitrary units) of pixel number $i$, $V$ is the 
average of $PV$ 
for the unexposed part of the plate, $B$ is the average of $PV$ for the  
darkest pixels, and $T_i$ is the $PV$ for pixel number $i$.  
The pixels of the unexposed part of the plate  are identified in each image as 
the ones outside the solar disk
with $PV$ higher than the maximum $PV$ of solar disk pixels. 
The darkest pixels in each un-calibrated image 
are the  ones with
$PV$ lower than or equal to the minimum $PV$ of solar disk pixels.
The effects of modifying the criteria for the identification of 
darkest and unexposed pixels 
upon the contents of the analyzed data are presented in Sect.~4.
 Note that the application of this straightforward 
 method for the photographic calibration of historic data is functional to our study, which takes into account only average image contents.  
The  accuracy of the applied method 
is also discussed in Sect.~4. 


The pixel values 
in the {\it Me} and {\it PSPT} images were provided by a CCD recording 
device. Given a proper instrumental calibration, the pixel values in both {\it Me} and {\it PSPT} images
are almost linearly proportional to the flux of radiation incident during the exposure  of the device. 


\subsection{Data inspection}
\begin{figure} 
\centering{
\includegraphics[width=4.25cm]{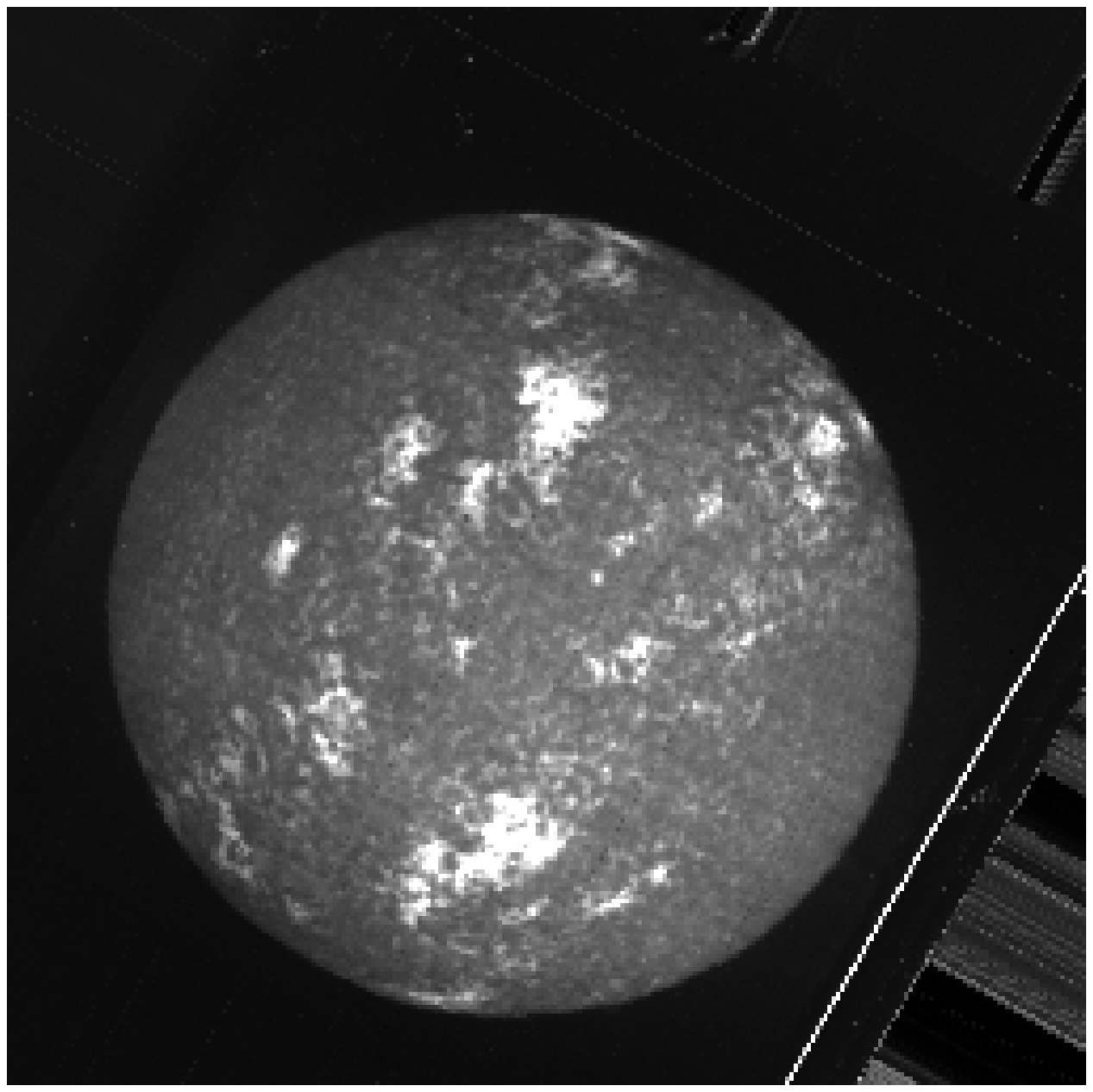}\includegraphics[width=4.25cm]{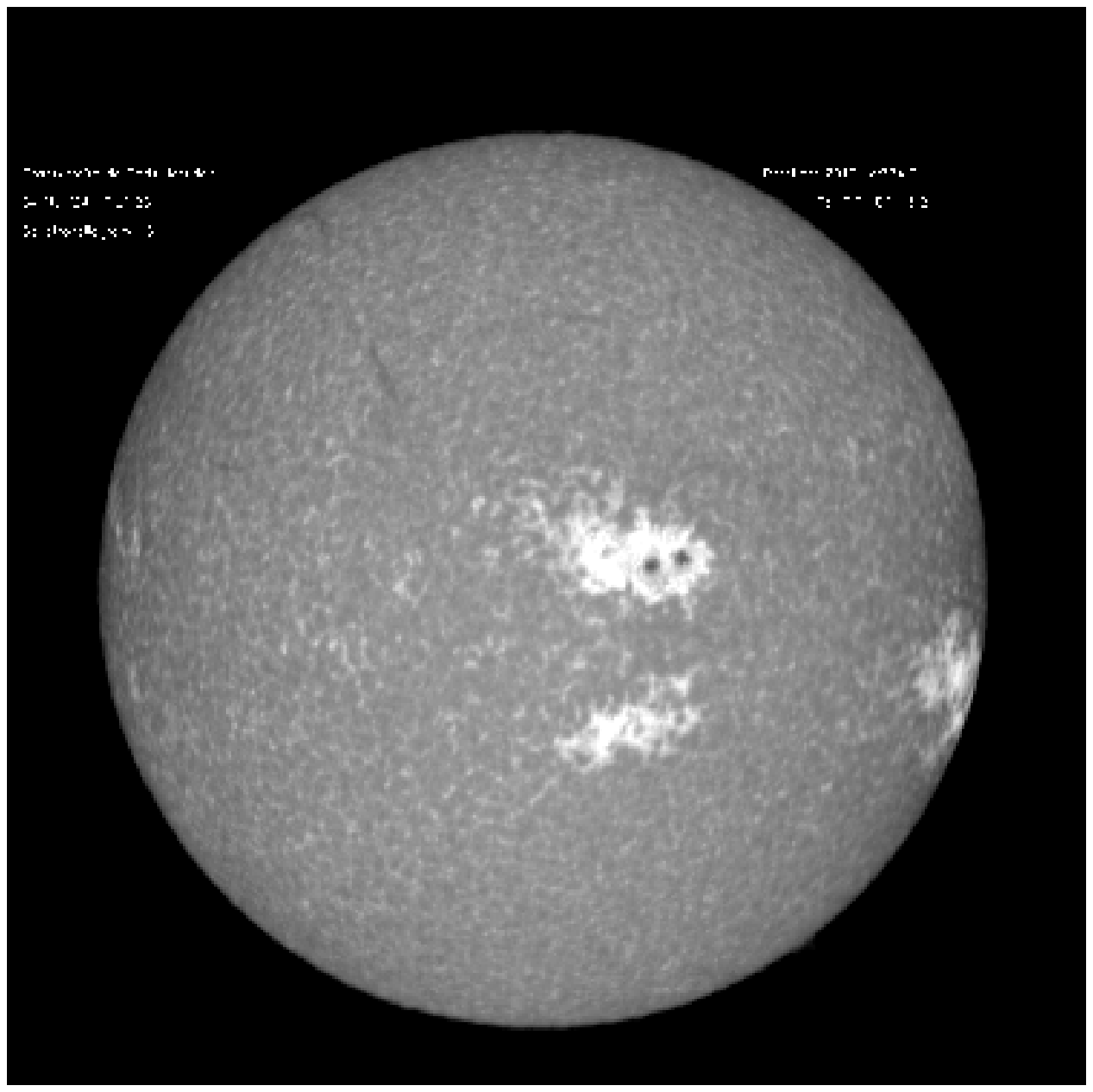}
\includegraphics[width=4.25cm]{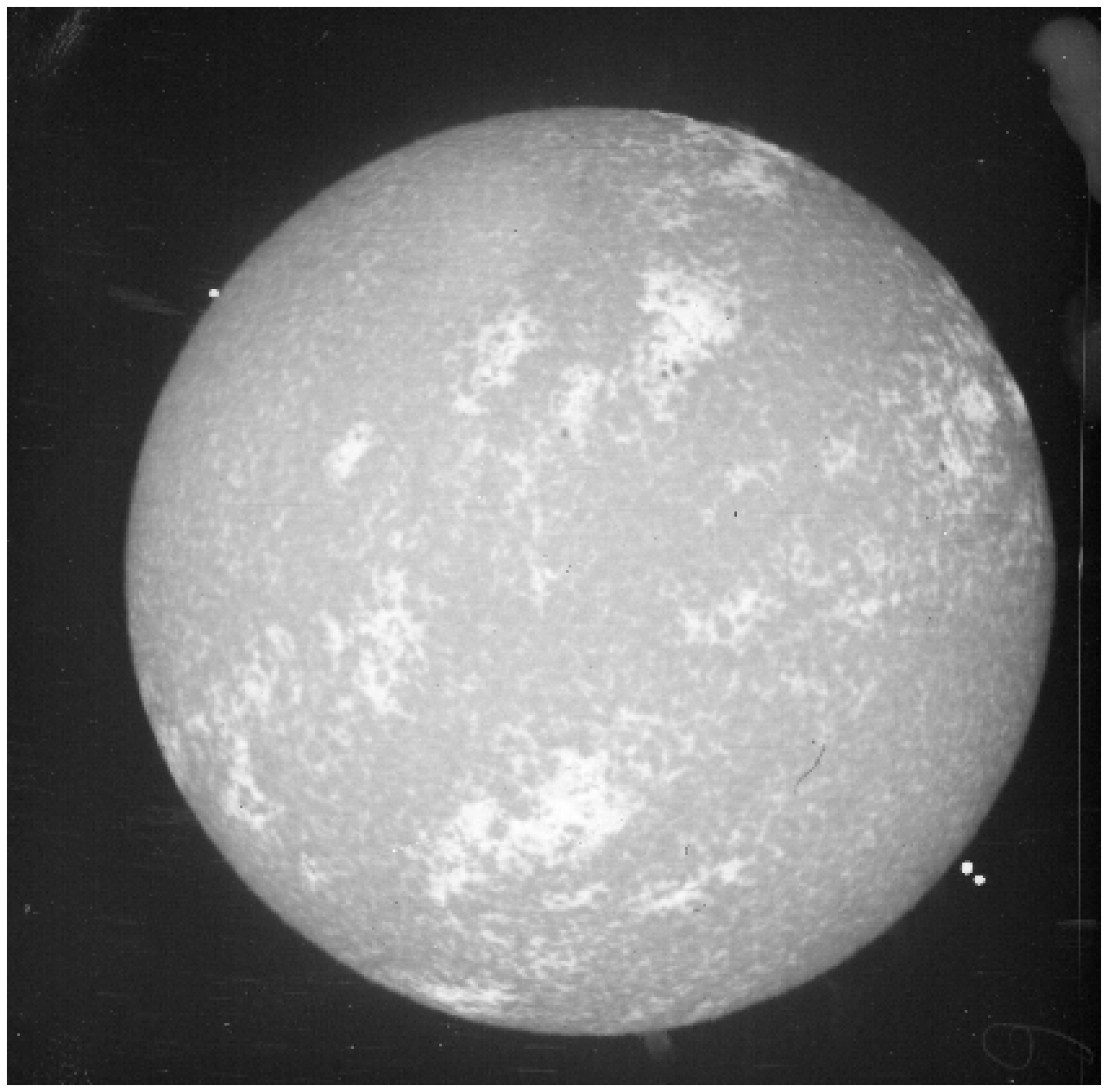}\includegraphics[width=4.25cm]{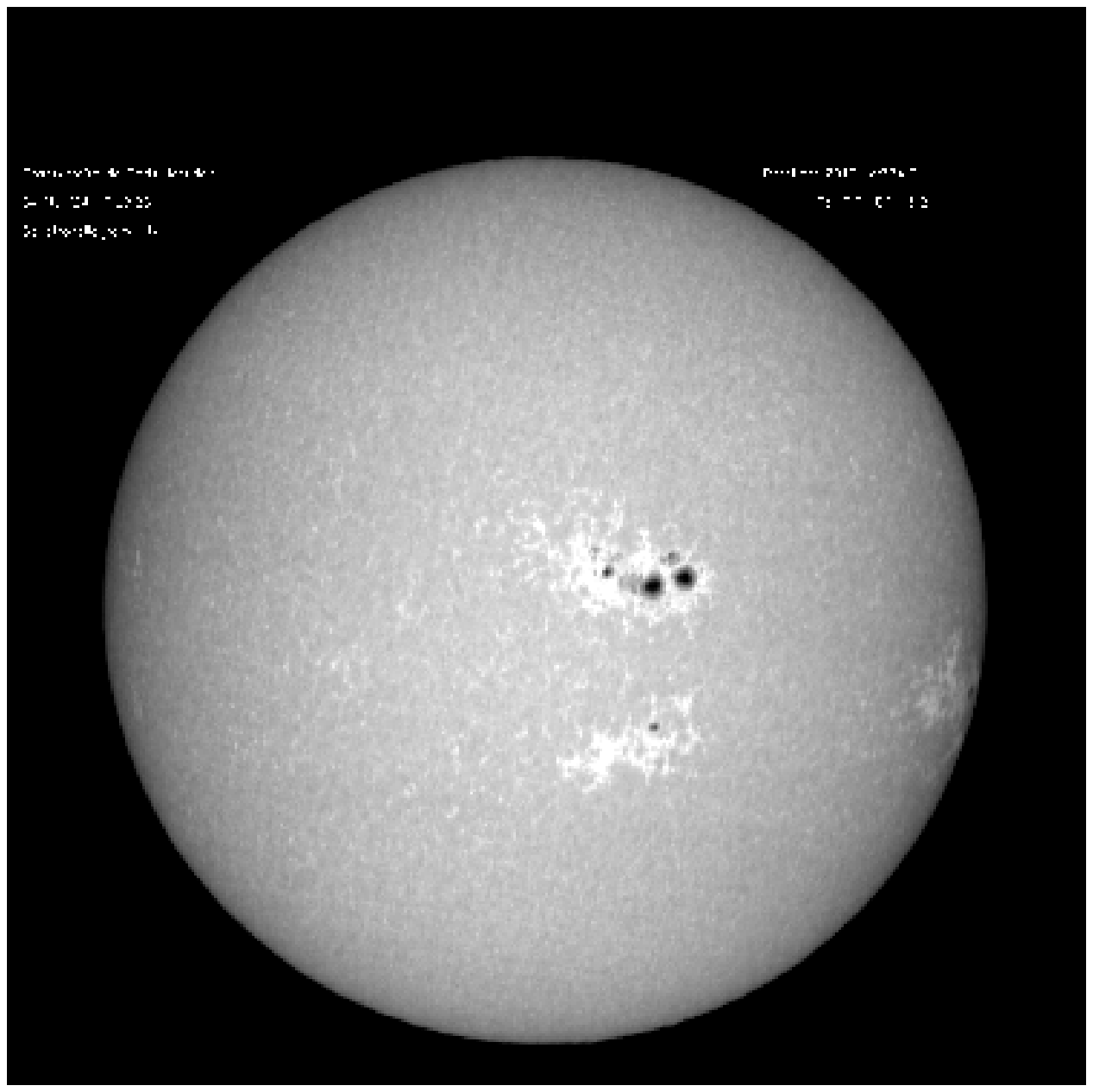}
\includegraphics[width=4.25cm]{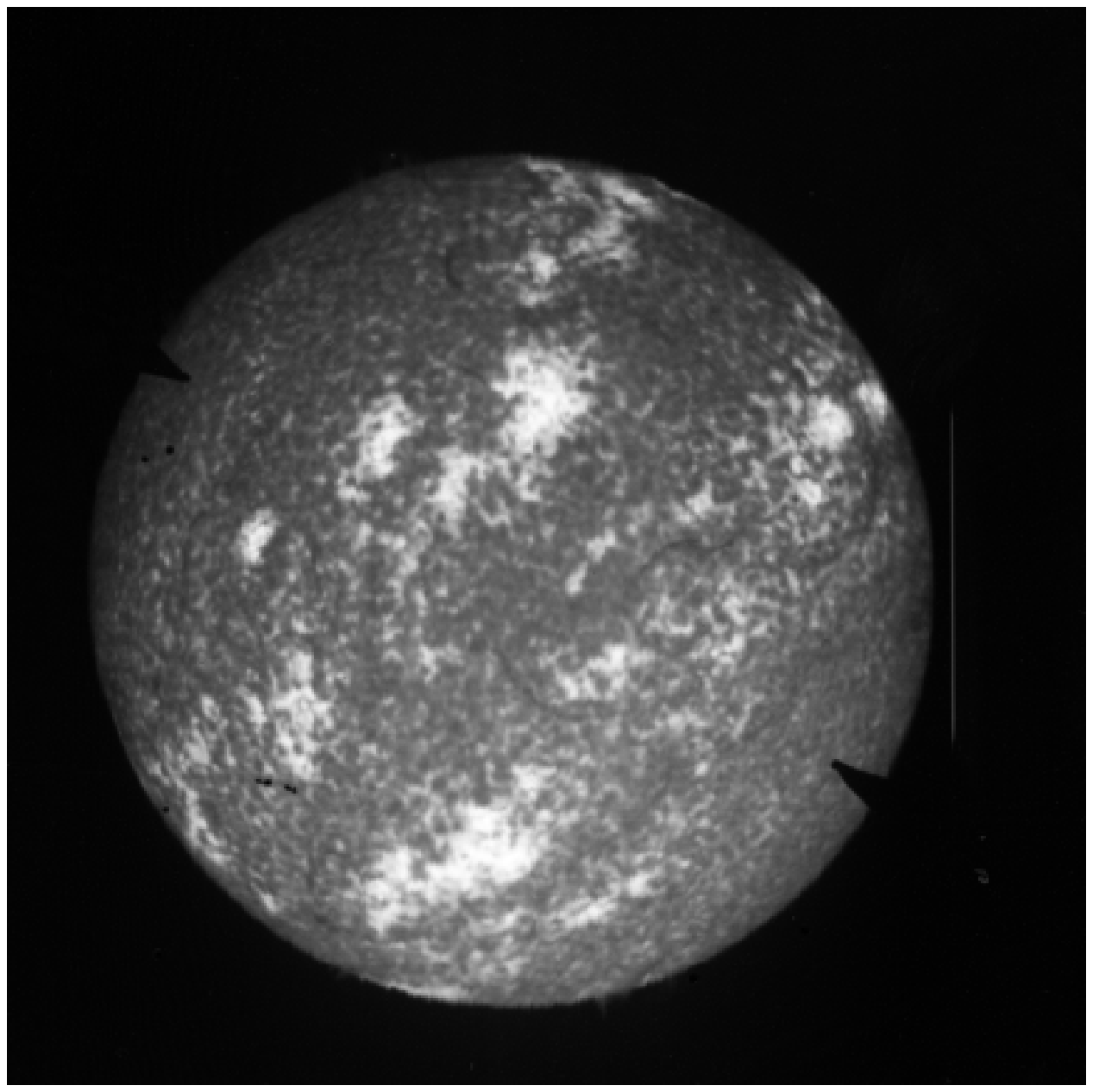}\includegraphics[width=4.25cm]{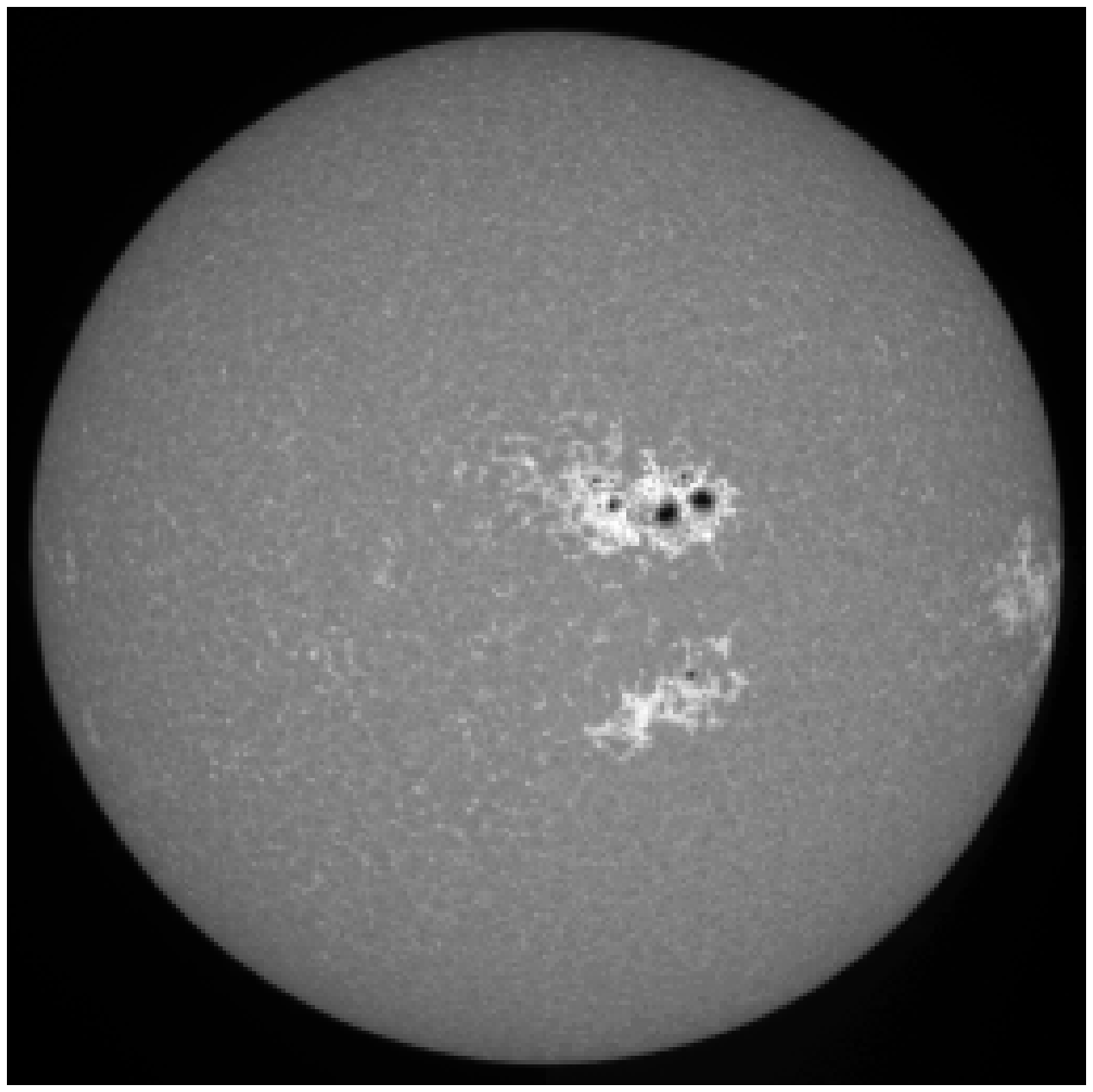}}
\caption{Examples of the Ca~II~K  observations analyzed in this study. Left: Arcetri (top), Kodaikanal (middle), and Mt Wilson (bottom) 
historic images obtained by the digitization of spectroheliograms taken at the three sites on 
January 9, 1958.  
All these images show sunspots when inspected at full resolution; the Mt Wilson image (bottom) clearly shows also on-disk filaments.
The Arcetri image (top) shows signs of a filament near the right limb (in the lower activity belt). Note that the 
pixel values were reversed, in order to show the brightness pattern as is usually  observed in the 
intensity images of the 
solar disk. Arcetri and Kodaikanal images were also rotated to liken the Mt Wilson solar observation.  
%
Right: $K_3$ (top) and $K_{1{\rm V}}$ (middle) Meudon spectroheliograms and Rome-{\it PSPT} filtergram (bottom) taken at the two sites on  July 24, 2004 
($\approx$~07:20~UT).}
\label{fig1} 
\end{figure} 
%

The visual inspection of images shows considerable differences between  the solar observations taken on the same day from the three historic 
series (Fig.~\ref{fig1}). 
For instance, the {\it MW} images show filaments over the solar disk, which are not found in the other data, although some are hinted at in the {\it Ar} images. On the other hand,  {\it Ko} images 
show sunspot regions which are almost absent in the {\it MW} observations, and  are partly found in the {\it Ar} data. Moreover, 
the position, dimension, and number of bright features identified in the images are quite different, 
especially close to the solar limb. 
The different observing time at the three sites, which may differ by more than 16 hours due to the site location, can explain part of these 
differences. However, the bulk of them arise from the 
different spectral sampling of these observations.
In particular, observations taken with a narrow spectral sampling centered at  $K_3$ show on-disk filaments, while the ones taken 
in a spectral range 
including the $K_{1{\rm V},1{\rm R}}$  show sunspots. 
  The {\it Me} and {\it PSPT}  data show all the solar features listed above. Their visual inspection also reveals small differences between 
  the observations
 taken at the two sites on the same day. 

\begin{figure} 
\centering{\includegraphics[width=8.5cm]{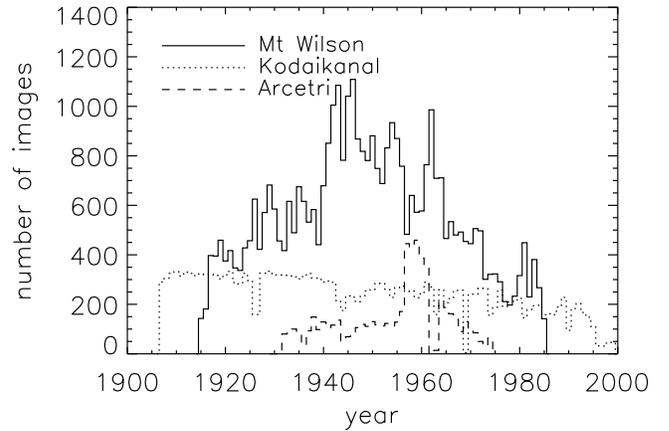}}
\caption{Temporal coverage of the Mt Wilson (solid line), Kodaikanal (dotted line) and Arcetri (dashed line) historic time-series and the number of 
Ca~II~K images per year analyzed in this study.}
\label{fig2} 
\end{figure}

The temporal coverage by the  historic series and the number of  
images per year available for our study are shown in Fig.~\ref{fig2}. It is seen  
that the {\it Ar} series contains a rather small number of images with respect 
to the other ones. However, more than 65\% of the images in this series
have exposure wedges for the calibration of the non-linear response of
the photographic emulsion.
In contrast, most of both {\it MW} and {\it Ko} images 
(75\% and 55\%, respectively) 
do not have wedge calibrations.
The {\it Ar} series is thus particularly suitable for a study and comparison of different calibration methods based on either calibration exposures or other criteria, such 
as the use of the solar intensity 
limb darkening or intensity distributions. 
The {\it Ko} series covers the largest period of observations. Since not all the original plates were digitized, 
the full set of data stored in the Kodaikanal archive is also more extensive than suggested by Fig.~\ref{fig2}. 
 However, the {\it Ko}  images  were digitized with a lower 
bit-significance compared 
to  the two other historic series. For instance, visual inspection of images reveals compression effects on the small scale blackening pattern, which are typically associated with lower resolution image 
 formats. Moreover, the 8-bit dynamic range of the digitized images often compresses the actual dynamic range of the blackening distribution on the original plates, mostly when 
plate defects, such as emulsion holes and scratches, are present. Because of this compression, the small-scale density patterns inside both darker and brighter 
features on the plates are lost.
Finally, the {\it MW} data look most promising, when taking into account both their nominally higher image quality and 
 the large number of images available, although over a shorter length of time
than the one covered by the {\it Ko} data. 

The visual inspection of the three historic series reveals some specific defects in these data caused by  
instrumental problems, as well as by the circumstances surrounding observations and storage. 
For instance, sometimes thick bright lines and bands in the direction of the spectrograph slits
are seen in un-calibrated images. 
These lines and bands
were likely caused either by
irregularities in the drive rate for the image scanning or by shadows due to the passage of clouds or due to objects in the optical path   
during observations. Sometimes variations of the resolution 
over the solar disk within one image are observed, which are caused by  changes in the 
atmospheric turbulence  during the observation. 
A number of un-calibrated images show tiny bright lines across the solar disk, perpendicular to the slit. These lines were probably produced 
by   irregularities or dust particles in the optical path of the spectroheliograph (e.g. on either entrance or exit slits). 
 On the other hand,  handling and
storage of the plates for  several decades  have also led damages in the plate emulsion. These are seen as bright scratches and holes in un-calibrated images.

Our  inspection has also revealed other frequent problems of these data. Variation of the solar declination during observation, 
curvature of the spectrograph slits, and small differences in the 
 velocity setting of drive motors in the instrument 
  resulted in geometric
  distortions of the solar image. 
   In particular, the solar disk often appears elliptic, the axes of the ellipse being parallel and 
  perpendicular to the spectrograph slit. 
Note that the position of these axes with respect to the horizontal and vertical directions in the analyzed images depends 
on the optical path of the spectroheliographs used to get the original photographic observations, as well as by the positioning of the 
photographic plate in the digitization device.  
 
 The historic un-calibrated data also display large-scale blackening patterns. Often they are most 
clearly seen outside the solar disk, but they also affect the solar images. These patterns are due to stray-light 
 introduced  by the
 instrument optics and by the turbulence in the Earth atmosphere. 
 Some images show localized blackening patterns, which may be due to chemical effects 
 during the plate development.
 Data inspection clearly shows also that blackening values 
in the historic solar images vary in a broad range. 
Such blackening differences are probably produced by
modifications of the instrumentation and of the observational procedures, as well as of the photographic
processing of the original plate. 
We also noticed occasional  spectrally  off-band observations. 

Note that the defects listed above occur to different degrees in the three analyzed historic series. 
For instance, the {\it Ko} images are, on average, less affected by defects, such as scratches, lines, 
and geometrical distortions, than both the {\it Ar} and {\it MW} data.
However, this is probably due to the fact that the {\it Ko} data-set
contains only the highest-quality plates, selected from the Kodaikanal archive prior to digitization.  

Finally, it is worth noting that the image size varies in two of the analyzed historic time-series, namely in {\it MW} and {\it Ko} sets. 
The size of {\it MW} images changes by up to 
about 4\%
over the 
whole series. It is typically larger for the data from October 1962 onward. This is mainly due to changes in the 
spectroheliograph on 8 October 1962. The new optics gave an about 1.75~times larger image of the Sun. In order to keep the size
of the digitized images more or less the same, a change in the scanning resolution was applied, which reduced the difference in the 
image size to less than 4\% but did not remedy it completely. 
The size of {\it Ko} images for three periods, namely 1927, 1950-1955, 
 and 1991-1999, 
is only half of the size of all  other {\it Ko} data.  Note that changes in the setup of the digitization device, such as the 
size of the digitized image, 
 may also determine 
variations of the image contents. {\it Ko} data during these periods are not included in our analysis. 
 This is meant to ensure that no bias due to the original different pixel sizes enters the results.

Most of the problems listed here, including variations of the image size, geometrical distortions, large scale
intensity patterns and stray-light, are discussed in more detail in Sect.~3.

\subsection{Image pre-processing}

The analyzed images of the {\it Ar} and the {\it MW} series were independently pre-processed in order to apply the flat-field calibration of the 
digitization device. Details about the applied methods are given by \citet{centrone2005} and by the UCLA project web-page for the {\it Ar} and {\it MW} 
series, respectively. 
On the other hand, visual inspection of {\it Ko} images shows that this series lacks such preliminary calibration. In fact, accumulation of dust and some defects 
in the digitization device can easily be noticed in the data.  
Due to the lack of information on the flat-field response of the digitization device, this pre-processing step 
is missing for this series. 
Note that the measurement of the average image contents considered in this study
is only slightly affected  
by dust accumulation and device defects. However, these defects may lead to inaccurate determination of solar 
features during any subsequent image-processing. 
The analyzed images of the {\it Me} and the {\it PSPT} series were independently pre-processed in order to apply the flat-field calibration 
of the CCD recording device. Details about the applied methods are given in the pertinent references listed in Sect.~2.1. 

Photographic calibration of images described in Sect. 2.2, as well as analysis of image contents described below, requires the knowledge of both, position of the center and radius  of the observed solar disk. 
For the {\it MW} images, information about the solar disk center, disk horizontal and vertical radii, and the 
quality of their measurements is stored in the FITS headers of images. They were determined by the UCLA project scientists by applying a four-step algorithm, 
which was designed to be 
robust against a variety of image defects and of image contents. The algorithm is repeated until 
trial results no longer change. For properly exposed images, it is known that the algorithm tends to over-estimate the 
solar disk size compared with typically adopted values. Details about the algorithm  can be found on the project web-page. 

The headers of {\it Ar, Ko, Me}, and {\it PSPT} images do not contain information about solar disk centre and radius measurements.
We determined each independently using a  method consisting of three  steps. It is partly based on 
automated techniques described in the literature \citep[e.g.][]{walton_etal1998,denker_etal1999,zharkova_etal2003}.
In brief, the method searches for the solar rim, by marking the location of pixels in which two selection criteria are  simultaneously fulfilled. 
The two criteria are: 1)  
the value of the marked pixel is higher than a given threshold value, which is computed taking into account the mean value of pixels 
belonging to a sub-array centered at the baricenter of  pixel values of the image; 
2) the gradient of  pixel values along the analyzed half-line  finds a  maximum 
value at the marked pixel.
The  search is performed  
along 360 half-lines, taking into account all pixels belonging to each  analyzed half-line until the criteria are fulfilled. 
The half-lines originate in the baricenter of  pixel values of the image.  
 The position of the marked pixels, i.e. the solar rim is then used for the computation of the disk center and shape, 
which is performed by applying an ellipse-fitting algorithm.
On average, this method gives values of the horizontal and vertical radii of {\it MW} solar observations which 
are about 1.5 pixel smaller than those given 
in the file headers. This difference corresponds to less than  
0.5\% of the radius.  
Note that  each historic series analyzed in this study shows specific characteristics and artifacts such that  the  algorithm for centre 
and radius measurements needs considerable  modifications for application to each series.

Most of the results presented in the following were obtained by analyzing re-sized images. This is to ensure  a 
similar solar disk size in all series, and 
thus to allow their direct comparison. Moreover, this image re-sizing also helps to compensate for the geometrical distortions affecting
 some of the analyzed data. 
Both horizontal and vertical radii of the solar disk  are re-sized to 350~pixels in 
all images, i.e.  roughly to the size of the {\it MW} data. 
By this image re-sizing, the 
{\it Ar} and {\it Ko} data are re-sampled to about half the linear size of the original digital images. For comparison, results 
obtained by the analysis of full-size data are also presented below.

\section{Data contents}
We measured several quantities in each image of the analyzed series in order to evaluate the data contents and the homogeneity 
in time
of such data. Measurements for historic data were performed on photographically calibrated images, 
which were obtained from the original data as described in Sect.~2.2. 

\subsection{Geometrical distortions}
\begin{figure} 
\centering{
\includegraphics[width=8.5cm]{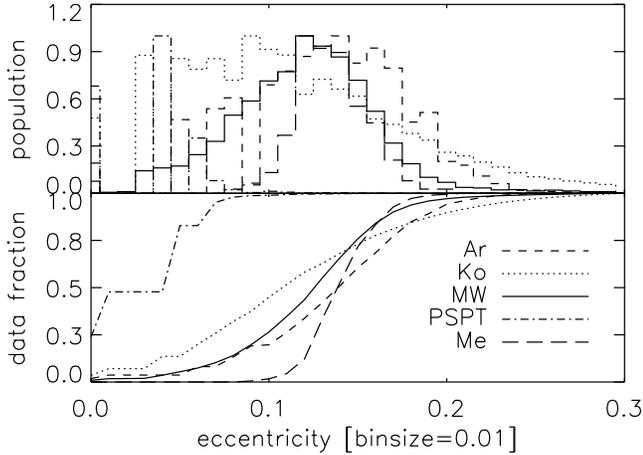}}
\caption{Solar disk eccentricity for the analyzed data.  
Top: Histograms of the eccentricity values. Bottom: Cumulative histograms 
of the measured values. The time-series corresponding to each line style is described by the legend in the bottom panel.}
\label{fig5} 
\end{figure} 
%
\begin{figure} 
\centering{
\includegraphics[width=8.5cm]{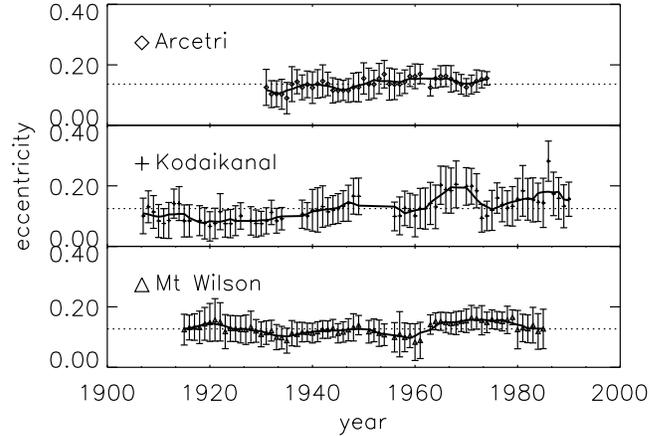}}
\caption{Temporal variation of the solar disk eccentricity for the {\it Ar, Ko}, and {\it MW} series. The error bars represent 
the dispersion of measurements
in terms of their standard deviation. The dotted line marks the median value of the average annual results for  the whole series.
The solid lines represent 5-year running means.}
\label{fig6} 
\end{figure} 
%
The main geometrical distortions affecting the analyzed data are revealed by the parameters of the ellipse which best fits the 
solar disk edge. We found that the  difference between the 
major and minor axes of the ellipse for the
{\it Ar},  {\it Ko}, and {\it MW} series is, on average, $\approx$ 2\%, $\le$1\%, and $\le$1\% of their mean value, 
respectively. 
The analysis  shows that {\it Ar} and {\it MW} series suffer from geometrical distortions, which are predominantly in the horizontal  and vertical 
directions.
In contrast, the orientation of the best-fit major and minor axes of the ellipse 
to {\it Ko} images rotates depending on the 
observation time. 
In order to compare homogeneous quantities for the data-sets, 
hereafter we take into account the ratio between the  horizontal $r_x$ and  vertical $r_y$ radii in each solar image. We also compute the solar 
disk
eccentricity $e = \sqrt{1-(r_{min}/r_{max})^2}$, where $r_{min},r_{max}$ are the smallest and the largest radii of the solar image in the horizontal and vertical
directions.
 
We found that the average value and the standard deviation of the disk eccentricity for the 
{\it Ar, Ko}, and {\it MW} images are  0.14$\pm$0.04, 0.12$\pm$0.06, and 0.12$\pm$0.04, respectively (Fig.~\ref{fig5}).
The same measurements carried out on {\it Me} and {\it PSPT} images give  
0.14$\pm$0.02 and 
0.04$\pm$0.03, respectively. We thus found that disk eccentricity of both historic ({\it  Ar, Ko, and MW}) and modern ({\it Me}) spectroheliogram 
observations is about 3 times 
larger than the one 
computed for modern observations 
taken with interference filters ({\it PSPT}). 

The standard deviation of annual averages of the disk eccentricity for the {\it Ar, Ko}, and {\it MW}
 time-series
is $\approx$ 0.02, 0.04, 0.02, respectively, i.e. about 14\%, 32\%, 15\% of the median values (Fig.~\ref{fig6}). However,  the 
standard deviation of 
the values measured over a year, due to, for example, seasonal variations and occasional 
 failures of the algorithms 
 used for radii measurements, is on average about twice as large as the variation of the  values on longer time-scales.
Note the slight continuous increase of disk eccentricity in the {\it Ar} series and the marked increase of the dispersion of results obtained 
for
the {\it Ko} data 
taken from about 1960 onward. This latter effect is mostly due to a marked decrease of the image quality and a subsequent increase of failures in the 
ellipse fitting calculations.  In contrast, the geometrical distortion of solar disk observations on {\it MW} data remains almost 
constant over the whole series. The measured value slightly increases for the data taken from 1962 onward.  
 

\subsection{Spatial resolution} 
\begin{figure} 
\centering{ 
\includegraphics[width=8.5cm]{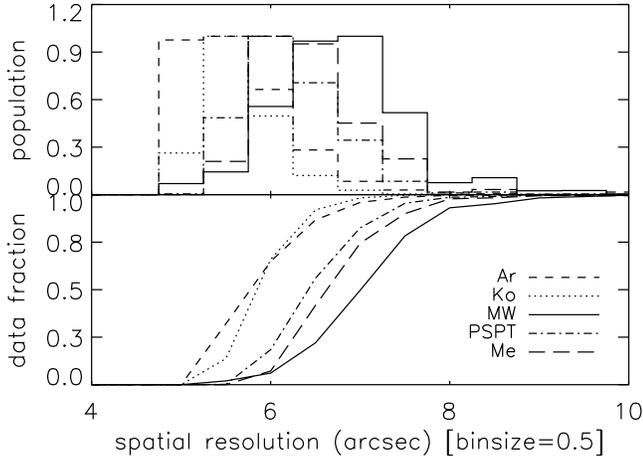}}
\caption{A measure of the spatial resolution  of  the analyzed data. 
Details are given in Sect.~3.2. Legend as in Fig.~\ref{fig5}.}
\label{fig7} 
\end{figure} 

\begin{figure} 
\centering{ 
\includegraphics[width=8.5cm]{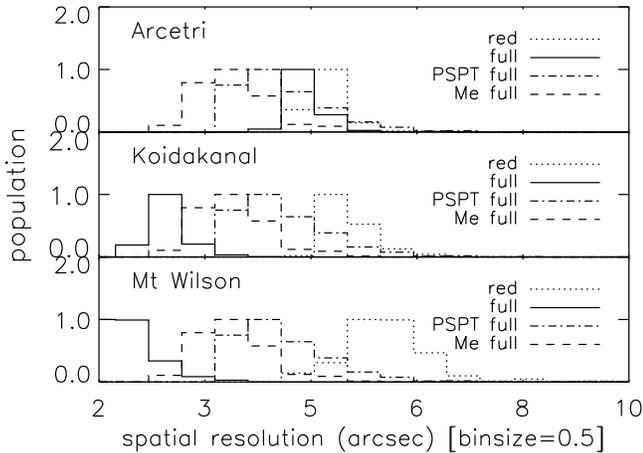}}
\caption{Histograms of the spatial resolution measured on both full-size (solid line) and reduced-size (dotted line) images corresponding to the same sample of original historic 
observations. 
The dashed and dot-dashed histograms show the results obtained for the full-size {\it Me} and {\it PSPT} data, respectively.}
\label{fig82} 
\end{figure} 
%

In order to evaluate the spatial resolution of each analyzed image, we studied the power-spectra of a 64$\times$64 sub-array extracted at 
solar disk center.  In particular, we measured for each image the spatial frequency at which 98\% of the image power spectral 
density is 
taken into account. 
The bulk of the information about patterns 
in the images lies at frequencies below this. 
The spatial scale 
corresponding to this measured cut-off frequency is taken as a measure of the spatial resolution in the analyzed image.
Note that the power density never decreases below a noise level for some data due to image defects. This is the reason we did not employ the 
usual technique of noting the spatial frequency at which the power drops to the noise level. The usage of a small 
sub-array and of the 98\% threshold is aimed to avoid the detection of power 
associated with the occurrence of solar features and image defects, respectively. 

The average spatial resolution  and its  standard deviation  on {\it Ar, Ko}, and {\it MW} images
are  
5.9$\pm$0.2$\arcsec$, 5.9$\pm$0.1$\arcsec$, and 6.9$\pm$0.6$\arcsec$, respectively, when considering images re-sized to the same solar disk size  
(Fig.~\ref{fig7}).
Thus, on average, {\it Ar} and
{\it Ko} images appear to carry more spatial information than the {\it MW} data, although the difference is rather small. 
On the other hand, the average spatial resolution of {\it MW} images is close to the ones of current 
{\it Me} and {\it PSPT} observations, which are  6.6$\pm$0.4$\arcsec$  and 6.5$\pm$0.3$\arcsec$, respectively. 
These results can be partly explained 
by the differences  in the  
 spectral sampling of the analyzed series. In fact, the nominal narrower spectral sampling of   {\it MW} data with respect to the other historic series, 
 corresponds to observations of higher atmospheric levels compared  to the others, and thus to observed features characterized by 
 lower spatial details.  This  holds also for {\it Me} images.

\begin{figure} 
\centering{ 
\includegraphics[width=8.5cm]{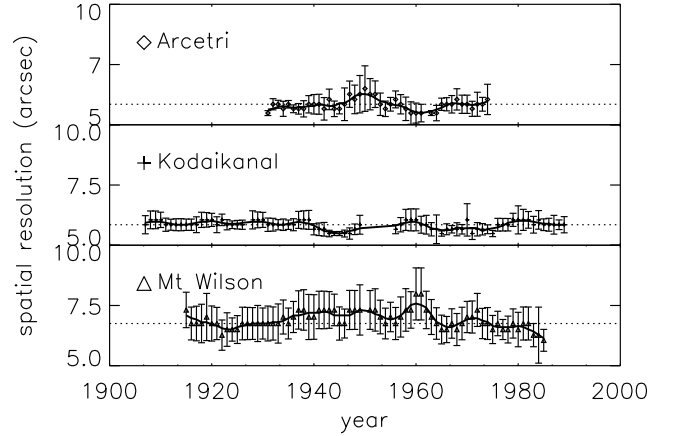}}
\caption{Temporal variation of the spatial resolution evaluated on the {\it Ar}, {\it Ko}, and {\it MW} images. Legend as in Fig.~\ref{fig6}.}
\label{fig8} 
\end{figure} 
 The
average spatial 
resolution computed for the historic data is  close to the one associated with the Nyquist frequency of the analyzed reduced-size images.  
In order to evaluate  whether the  image re-sizing leads to a  loss of spatial resolution stored in the original data, 
we 
analyzed samples of {\it Ar, Ko}, and {\it MW} full-size images. We found that the average spatial resolution and its standard deviation 
in the sample of full-size {\it MW} images  are 2.6$\pm$0.2$\arcsec$, compared to 6.6$\pm$0.5$\arcsec$  obtained for the same sample of observations 
with the  
reduced-size images (Fig.~\ref{fig82}).  These values for 
{\it Ko} full-size data are  3.3$\pm$0.1$\arcsec$, compared to  6.04$\pm$0.13$\arcsec$, for the same sample of observations with the reduced-size images. 
Finally, the values obtained for 
{\it Ar} full-size data are  5.4$\pm$0.1$\arcsec$, compared to  5.74$\pm$0.06$\arcsec$, for the same sample of observations with the reduced-size images. 
For comparison, the same quantities evaluated in the samples of full-size
{\it Me} and {\it PSPT} images are 5.0$\pm$5.0$\arcsec$ and 5.0$\pm$0.4$\arcsec$, respectively. 

The average spatial resolution measured    on {\it Ar}, {\it Ko}, and {\it MW} 
full-size images is also close to the one limited by the spatial sampling of the analyzed data. Moreover, we found that 
the values of the
average resolution measured at the solar disk centre and outside the solar disk are within 
one standard deviation in measurements.  
These results suggest that  the power density found at the smallest spatial scales in the 
full-size data depends on image digitization. 
We then analyzed the resolution measured at the solar disk centre  on images re-sized  using linear interpolation to 
half, one-third,  
and one-fourth the original dimension. 
We 
found that 
the resolution measured on each set of re-sized images 
shows  a skewed distribution of values, 
with a longer tails to the right of the distribution maximum. 
We computed the moments of these distributions. The obtained values 
indicate that the distribution resulting from 
images re-sized to half the original dimension
is the most symmetric about the maximum, as well as the less peaked, among the ones we studied. The obtained values 
also suggest that the 
re-sizing of images to more than half the original dimension implies a loss of spatial information for the solar observations  
stored in the full-size data.



 The analysis of the temporal variation of the image resolution for each series allows an evaluation of the  
 homogeneity of the data in each data-set. We found that  the spatial 
 scale of all the three historic series  varies slightly in time  (Fig.~\ref{fig8}). In particular, 
 the standard deviation of averaged annual values for the {\it Ar, Ko}, and {\it MW} series is about  4\%, 3\%, 5\%, respectively. However,
 the standard deviation of values measured over a year  
 is on average about twice larger, due 
 to seasonal variations of the image contents.  
 
Figure~\ref{fig8} shows that the  spatial resolution on the {\it MW} images has steadily increased in the last two decades of observations. 
This variation is likely due to   
instrumental changes (e.g. differences in the spectral sampling of the data). 
Note that the variation of image size affecting the {\it MW} data also contributes to the increase of the spatial resolution measured   
for {\it MW} images from 1960 onward. However, the measured resolution increase  is about 20\% larger than the one expected by taking into account the  
variation of the solar disk size in the  analyzed images. 
Finally, our results suggest that the spatial resolution of the {\it Ko} data remains almost constant over the whole period.

\subsection{Large-scale image inhomogeneities} 
\begin{figure} 
\centering{ 
\includegraphics[width=8.5cm]{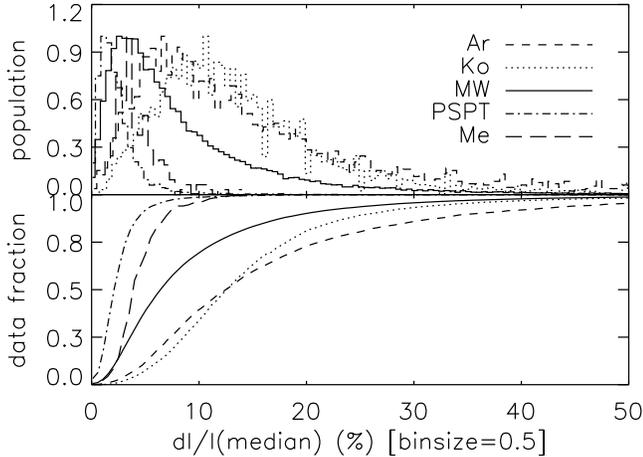}}
\caption{A measure of the large-scale inhomogeneities affecting the analyzed data. 
Details are given in Sect.~3.3. Legend as in Fig.~\ref{fig5}.}
\label{fig11} 
\end{figure} 
\begin{figure} 
\centering{ 
\includegraphics[width=8.5cm]{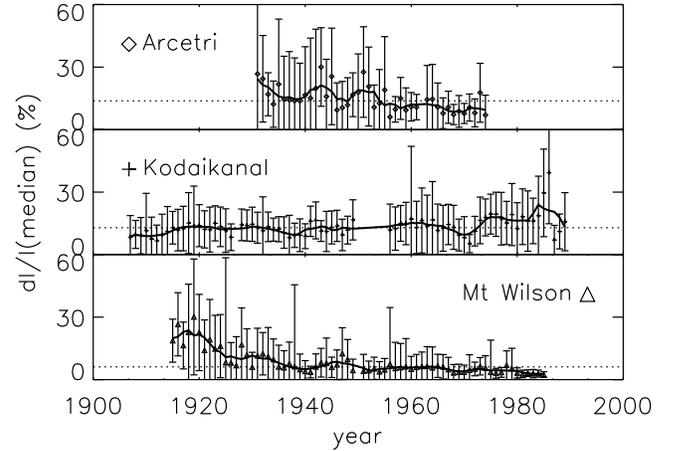}}
\caption{Temporal variation of the results obtained measuring the large-scale inhomogeneities affecting the {\it Ar, Ko}, and {\it MW} images.
Legend as in Fig.~\ref{fig6}.}
\label{fig12} 
\end{figure} 
Next we considered some photometric properties of the analyzed series. 
We first evaluated the level of  large-scale 
inhomogeneities in  the  
images introduced by  variations of the sky transparency during observations and by instrumental problems. In particular, we analyzed 
the median intensity over a   
ring centered on the solar disk center of each image. The ring spans the disk positions  $\mu$=0.50$\pm$0.05.
Since strong inhomogeneities may affect only a portion of the solar image, we divided the solar disk into four quadrants and 
 calculated the deviation of the 
median intensity value in each quadrant, with respect to the median value over the whole ring. We then took the maximum deviation  
measured over the four quadrants as a measure 
of the degree of large-scale intensity inhomogeneities  affecting the solar disk images.  Note that usage of  median intensity values and of the  
disk positions $\mu$=0.50$\pm$0.05 is aimed to lower the influence of active regions on the obtained results. 


We found that the median value and the standard deviation of measurements for  the {\it Ar, Ko}, and {\it MW} series are
13$\pm$17\%, 12$\pm$6\%, and 6$\pm$6\%, respectively (Fig.~\ref{fig11}). The same quantities evaluated for the 
{\it Me} and {\it PSPT} data are  3.9$\pm$2.2\% and 2.2$\pm$2.2\%, respectively.   
The standard deviation of measurements for each series is about 43\%, 37\%, and 96\%, which is close to the 
average of the standard deviation  of values measured over a year for the {\it Ar} and {\it MW} series, and half the one for the {\it Ko} series. 
 
We found 
that the  first half of the {\it Ar} series, as well {\it MW} data taken during the first decade of observation 
 suffer from strong large-scale inhomogeneities (Fig.~\ref{fig12}). 
This is in agreement with information stored in {\it MW} log-book notes which report that  image vignetting was very strong at that time 
due to the use of an under-sized grating. 

\subsection{Stray-light}
\begin{figure} 
\centering{
\includegraphics[width=8.5cm]{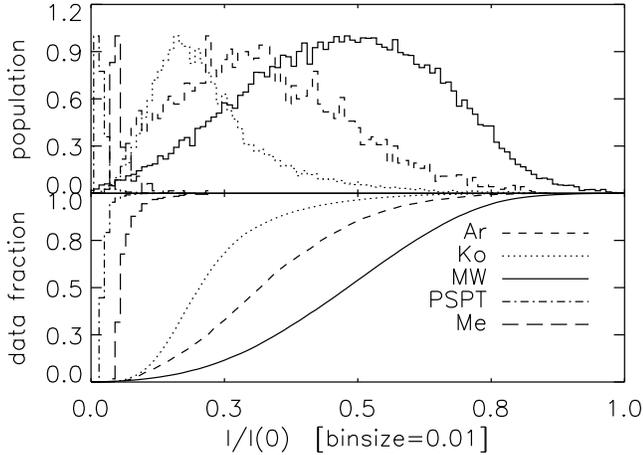}}
\caption{A measure of stray-light level on the  analyzed data. 
These results concern aureola measurements at $r/R_{Sun}=1.06\pm0.01$. Details are given in Sect.~3.4. Legend as in Fig.~\ref{fig5}.}
\label{fig13} 
\end{figure} 
\begin{figure} 
\centering{ 
\includegraphics[width=8.5cm]{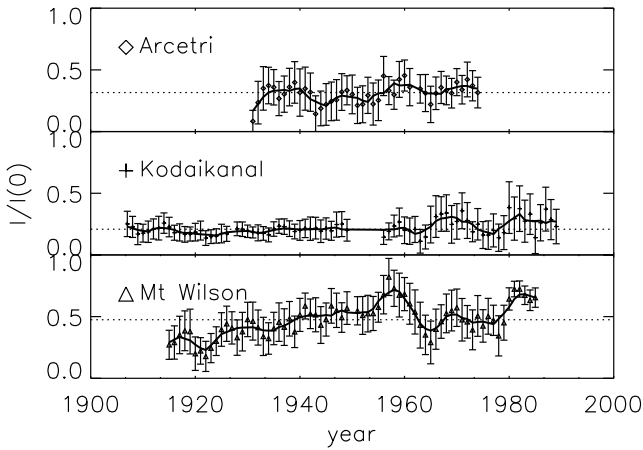}}
\caption{Temporal variation of  stray-light level on the {\it Ar, Ko}, and {\it MW} images. 
Legend as in Fig.~\ref{fig6}.}
\label{fig14} 
\end{figure} 

We  evaluated 
the level of stray-light  in the data by checking  the fall-off of intensity values just outside the 
solar limb (aureola, hereafter) in each image. For this, we  computed the median intensity over rings centered on $\mu=1$
and evaluated 
 the profile  of these median values  in each image, starting from the disk centre up 
to the image edge. 
Each ring is one pixel thick.  
In order to compare the stray-light level in all series, we have then used these calculated radial profiles 
to evaluate  the ratio
of  the intensity values at fixed off-limb distances (e.g., $r/R_{sun}$=1.06$\pm$0.01, 1.125$\pm$0.025, 1.225$\pm$0.025) to that at the disk center 
($I/I(0)$, hereafter).  Note that analysis of aureola regions  close 
to the solar disk ($r/R_{sun}<1.3$) is aimed to avoid the influence of strong intensity inhomogeneities often observed outside of the solar disk. These
inhomogeneities  are mostly due to unexposed 
plate regions, calibration exposures, and inscriptions on the original plate.
Figure \ref{fig13} compares the results of our estimates of the stray-light at $r/R_{sun}$=1.06$\pm$0.01  
for all data-sets calibrated as described in Sect.~2.2. 
The dependence of the obtained results on the  calibration method applied to the original data  
is presented in Sect.~4.  

We found that historic  {\it Ar, Ko}, and {\it MW}  images 
 are strongly affected by stray-light, much more than modern {\it Me} and {\it PSPT} data. Moreover,    
{\it Me} spectroheliograms suffer more from stray-light than  {\it PSPT} filtergrams.
The median value 
 and the standard deviation
of the measured intensity ratio at  $r/R_{sun}$=1.06$\pm$0.01  for {\it Ar, Ko}, and {\it MW} images are
0.32$\pm$0.15, 0.20$\pm$0.11, and 0.48$\pm$0.18, respectively. The same
quantities for {\it Me} and {\it PSPT} observations are
 0.05$\pm$0.03 and 0.02$\pm$0.01, respectively.
 
The temporal variation of the stray-light level is sensitive to changes in instrumental conditions and setups.  
For instance, Fig.~\ref{fig14} shows a large reduction of the stray-light level 
for the {\it MW} data taken from 1960 onward,  probably due to the
installation of new gratings, which is also recorded in the observatory log-books. 
The results also clearly show effects of the aging of both instruments  and 
observing programs. In fact, the stray-light level increased with grating use, since that component degraded in
the open air installation of the spectrograph.  The standard deviation of annual averages over the {\it Ar, Ko}, and {\it MW} series
is about 25\%, 29\%, and 28\%, respectively. 
 The standard deviation of  values measured over a year due 
 to seasonal variations of the image quality
 is about 8, 1.3, and 1 times larger for the three series, respectively.  

\subsection{Image contrast} 
\begin{figure} 
\centering{ 
\includegraphics[width=8.5cm]{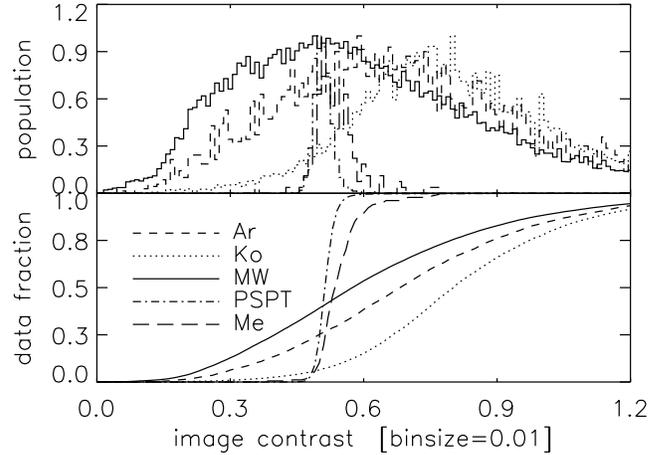}}
\caption{A measure of the image contrast for the  analyzed data.  Details are given in Sect.~3.5. Legend as in Fig.~\ref{fig5}.}
\label{fig9} 
\end{figure} 
\begin{figure} 
\centering{ 
\includegraphics[width=8.5cm]{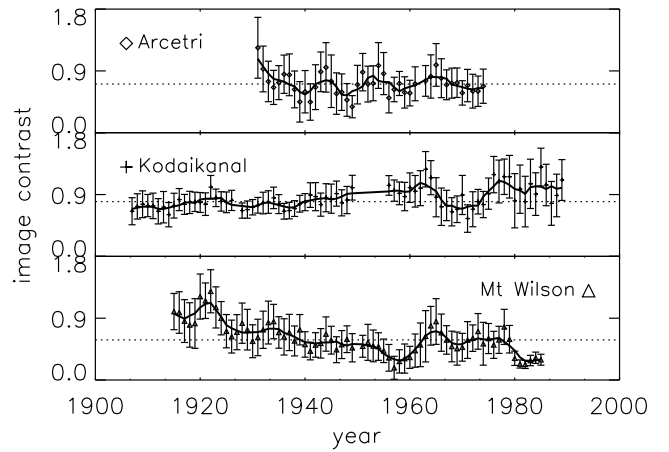}}
\caption{Temporal variation of the image contrast on the {\it Ar}, {\it Ko}, and {\it MW} images.
Legend as in Fig.~\ref{fig6}.}
\label{fig10} 
\end{figure} 
Next we considered the range of intensities measured in the solar images of each time-series. 
In particular, we analyzed the center-to-limb variation  of 
intensity values of quiet Sun regions ({\it CLV}, hereafter) and measured 
the maximum minus the minimum intensity values, normalized to their mean  
(henceforth, image contrast). The {\it CLV} was evaluated  on each image 
by computing the median of intensity values over each of 20 rings of equal area centered on the solar disk center. Maximum and minimum intensity 
values are the largest and the smallest values among the 20 intensity values thus determined. Note that usage of the small number of rings and of the 
median intensity in each ring is aimed to lower the influence of image defects and of active regions on the obtained results. 

The median value and the standard deviation  of the image contrast for  
{\it Ar, Ko}, and {\it MW} images are 0.7$\pm$0.3, 0.8$\pm$0.2, and 0.6$\pm$0.3, respectively (Fig.~\ref{fig9}). The same quantities  
for {\it Me} and {\it PSPT} images are   0.53$\pm$0.05 and 0.51$\pm$0.04, respectively.
 Note that the standard deviation of contrast measurements for modern data is 
about 10 times lower than that for historic data.  This may partly  be due to  the larger samples of images covering longer 
periods of time compared to the modern sets. However, the main reason is probably the higher 
degradation due to stray-light and the uncertainties in the photographic calibration of the  historic data. 
For instance, the method applied in this study to perform the photographic calibration makes use of pixel values of both
unexposed and dark regions of the original plate, which might be  badly defined in the analyzed data. 
Note that 
the median value and the standard deviation  of measured image contrasts
for the three historic series are very similar. Thus the results obtained are not affected by differences in the solar disk center determination and shape 
calculations performed on the 
three series.  




Figure \ref{fig10} shows the temporal variation of image contrast for the three historic time-series.
Changes in instrumentation and the observing procedure are well seen. 
The annual averages of the image contrast is found to
vary by about 0.17, 0.16, and 0.23, i.e. by about 23\%, 20\%, and 40\% of the median values for the {\it Ar, Ko}, and {\it MW}
 series, respectively. The standard deviation of values measured over a year, which takes into account seasonal variations of the image quality, 
 instrumental changes, 
 and  occasional failures of the algorithms 
 used for radii measurements, is close to the standard deviation of the annual averages.  

\section{Discussion}
We have analyzed the image contents of three historic time-series of 
Ca~II~K observations obtained by the digitization of photographic archives of
Arcetri, Kodaikanal, and Mt Wilson  spectroheliograms. 
The results have been compared to those obtained for the modern   
synoptic Ca~II~K observations taken with the Meudon 
spectroheliograph and the 
Rome-PSPT telescope. We have also  analyzed the temporal variation of the image contents, in order to evaluate the homogeneity of the three 
historic time-series.
 
This study shows that historic spectroheliograms suffer from stronger geometrical degradation and large-scale inhomogeneities 
than modern filtergrams, but at a similar level to present-day spectroheliograms.  
 Historic data also suffer from stronger pixel-scale image defects  than current observations.
Some of these defects are accountable to the aging of the original observations, some are due to shortcomings of the instruments. 
They can partly be corrected for or taken into account by a proper analysis. We find these issues bothersome, but not as serious as problems 
related to the photometric properties of the historic spectroheliograms. 
In particular, both the average value and the standard deviation of the stray-light level are  higher in the historic data than in the 
modern ones.  
 This holds also for the image contrast.
The difference in the image contrast is particularly worrisome since this lies at the heart of the scientific evaluation of the historic images. 
It may be caused by 
differences in spectral sampling (e.g. if historic and modern data were to sample different parts of the line profile), calibration issues, or 
degradations due to stray-light.

Peculiarities in the spectral sampling of the historic data  seem unlikely to be responsible for the differences in image contrast described above. 
In fact, we measured the relative difference between the {\it CLV} curves computed for  $K_3$ and $K_{1{\rm V}}$  {\it Me} images recorded 
on average less than 2 minutes apart. 
We found that 
the dispersion of  the {\it CLV} curves computed for 
the {\it Me} images obtained with the two spectral 
samplings in the same observing day is smaller than the dispersion of  results obtained for  spectrally homogeneous data taken on different days.
In particular, 
the median value and the standard deviation of 
the relative difference between the {\it CLV} curves computed for the two samples of 
spectral images are -0.006$\pm$0.016 for all disk positions $\mu\ge0.4$. 
The same quantities computed
for the sample of $K_3$ images, 
with respect to a randomly selected $K_3$  image 
are  -0.003$\pm$0.020. 
This  result is
 in agreement with the findings 
of a similar  analysis which we performed on a few  
 Ca II K spectroheliograms taken recently at the Coimbra Observatory\footnote{This set of  observations is composed of images obtained 
 for different spectral sampling by the instrument, both 
the spectral position and the bandwidth along the Ca~II~K line. The images were kindly provided by  A. Garcia (Coimbra Observatory).}.  

In order to evaluate the effects of the photographic calibration 
on the results presented above,  
we analyzed  images obtained by applying three different calibration methods to the same sample of 713 {\it MW}  
observations.  
The three methods are: 1) the one applied by the UCLA project scientists ({\it UCLA}, hereafter), which takes into account 
the calibrated 
exposures available on the side of the solar observations; 2) the method described in Sect.~2.2 ({\it calib}, hereafter); 3) the method which assumes
a linear relation between the  pixel values in the analyzed images and the incident flux  ({\it linear}, hereafter). 
It is worth noting that after the conversion of the plate blackening to intensity through the use of the step-wedge exposure, 
the {\it MW} images were further adjusted by the UCLA project scientists for the effect of a variable vignetting function, which 
yield 
large-scale intensity patterns in the observations  not associated with the solar limb-darkening variation of intensity.  
The vignetting function 
was derived by the UCLA
project scientists under the assumption that the limb-darkening variation of intensity computed on calibrated images  
matches some modern measurements \citep{livingston2008}.
After this adjustment, the {\it CLV} evaluated on each {\it MW} image 
depends on both the photographic calibration applied to original images and on the removal of vignetting effects on images. 
Therefore the value of photometric measurements performed on the data  relies on the precision of both image processing steps.

%

We found that the range of intensity 
values for quiet Sun regions obtained from the {\it UCLA} calibrated sample   is smaller
than that computed for the other two samples of images  for all disk positions $\mu>0.25$, which corresponds to  about 95\% of the solar disk.  
The  dispersion of CLV curves over 90\% of the solar disk is about 25\%, 33\%, and 76\% for the {\it linear}, {\it calib}, 
and 
{\it UCLA} image samples, 
respectively. In summary,  the intensity {\it CLV} of quiet Sun regions computed for the {\it UCLA} calibrated sample is flatter than the ones obtained 
with the other two samples of data, but 
the dispersion of curves computed from one day to the next for the {\it UCLA} sample is larger than the ones for the other samples. 

 
\begin{figure} 
\centering{
\includegraphics[width=8.5cm]{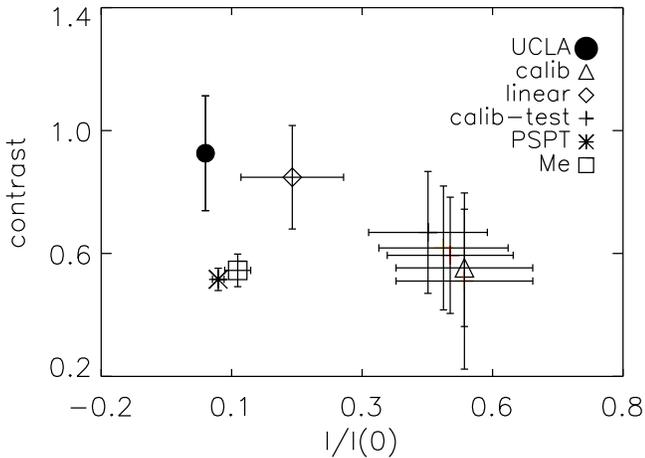}}
\caption{Scatter plot between the mean values of image contrast (ordinate, contrast) and stray-light measurements (abscissa, $I/I(0)$)
obtained for {\it MW} historic observations taken in 1967
and photographically calibrated with three different methods (UCLA, calib, and linear).
Results obtained by applying the calib method through different numerical criteria are also shown (calib-test).  Details are given in Sect.~4. 
For comparison, the same quantities evaluated for modern {\it Me} and {\it PSPT}
data are also plotted.
 Error bars represent the standard deviation of measurement results.}
\label{fig_scatter_c_auro} 
\end{figure} 
%

 
Figure \ref{fig_scatter_c_auro} shows 
 the relation  between image contrast and aureola intensity measurements for the three data-sets of images obtained by 
 applying the {\it UCLA}, {\it calib}, and {\it linear} calibration methods to the same set of {\it MW} observations. For comparison, results obtained for
modern data are also shown. 
We found that  
the mean value of the image
contrast measured on {\it MW} images varies up to about 25\% depending on the calibration method applied to the data. 
Figure \ref{fig_scatter_c_auro} shows also the results obtained by applying the {\it calib} method with  
different numerical criteria for 
the identification of the 
unexposed and the darkest pixels of the images.
In particular, the pixel identification was performed by using four sets of threshold values, which are  
based on: 1) maximum and minimum values of $PV$ measured
inside the solar disk;  2)
maximum and minimum values of $PV$ measured on the whole image; 3) higher than $m+3\sigma$ and lower than $m-3\sigma$, where $m$ and $\sigma$ are the 
mean and the standard deviation of 
 $PV$ measured 
inside the solar disk, respectively; 4) constant values (32767 and 0, respectively). The results  obtained   ({\it calib-test}, hereafter)  
are shown in Fig. \ref{fig_scatter_c_auro}, represented by crosses. We found that the image 
contrast measured on the analyzed data   
changes by about 10\%
by modifying the numerical criteria as described above.


Note that the  values measured on both
{\it calib} and {\it calib-test} images  approaches the ones resulted for modern observations ({\it Me} and {\it PSPT}). 
However, the standard deviation of the values measured 
for both {\it calib} and {\it calib-test} historic images is about 4 times larger  than the ones resulting for modern observations. 

\section{Conclusions}

We have analyzed the image contents of three historic time-series of Ca~II~K spectroheliograms obtained by the digitization of the Arcetri, 
Kodaikanal, and 
Mt Wilson photographic archives. Perhaps unsurprisingly, our study shows that historic data suffer from stronger degradation effects associated with instrumental problems than 
similar modern observations. Some of the image problems described in this study, e.g. geometrical and image defects, can be fixed through the 
development and 
application of  a proper image processing technique, as we have shown. For instance, the solar disk ellipticity can be compensated
by re-sizing the analyzed images. Some other problems, such as the  degradation of the spatial resolution within one image due, for example, to instrumental problems, 
can only partly be solved through the application of sophisticated image processing. Historic data also suffer from strong photometric 
 uncertainties due to often missing or poor photographic calibration. In addition, stray-light effects are much stronger in historic data than in modern observations.  
Therefore a special analysis is needed in order to check whether and to what extent the methods presented in the literature 
for the photometric calibration of data can restore the  historic Ca~II~K images to a homogeneous data-set with trustable intensities. 
The discussion of methods for the photographic calibration and the stray-light correction of these  images 
will be addressed in future papers.

Our results also show  that the image contents of the three considered historic data-sets vary in time. 
These variations are, to a large extent, due to changes in and aging of the instrumentation and evolution of the 
observing programs. 
The effects of the multiple instrumental changes over many decades are even more difficult to 
account for with image processing than the image defects mentioned above. 
The temporal variations of the image contents due to instrumental changes can be singled 
 out from solar temporal variations only through the inter-calibration of the data coming from different archives.
Our results   suggest 
that  for such inter-calibration it would be extremely useful to digitize the {\it Ko} series with a higher quality 
than available at present, since the 
{\it Ko} series turns out to be the most homogeneous and longest among those considered.

This study also shows that  the reliability of photometric measurements performed on historic data relies on the 
precision of their photographic calibration and on the removal of stray-light effects.  The main challenge for 
the analysis of such data is thus their accurate photometric 
calibration, for them to provide  value for studies concerning long-term solar activity and variability.


\acknowledgements
The authors thank the  Arcetri, Kodaikanal, Meudon, Mt Wilson and the Rome Solar Groups 
for the data provided. J. Aboudarham, F. Cavallini and J.M. Malherbe are acknowledged 
for useful discussions. 
This work was partly supported by the CVS project (Regione Lazio) and by the Deutsche Forschungsgemeinshaft, DFG project SO-711/1-2.
The digitization of the Mt Wilson Photographic Archive has been supported by the US National Science Foundation grant ATM/ST 0236682.


\begin{thebibliography}{aa}

\bibitem[Bappu(1967)]{bappu1967}
Bappu, M.~K.~V. 1967, \solphys, 1, 151

\bibitem[Centrone et al.(2005)]{centrone2005}
Centrone, M., Ermolli, I., \& Giorgi, F. 2005, Mem. SAIt. 76, 941

\bibitem[Chapman et al.(2004)]{chapman2004}
Chapman, G.~A., Cookson, A.~M., Dobias, J.~J., Preminger, D.~G., \& Walton, S.~R. 2004, Advances in Space Research, 34, 262

\bibitem[Coulter \& Kuhn(1994)]{coulter1994}
Coulter, R.~L. \& Kuhn, J.~F. 1994, 
"Solar active region evolution: comparing models with observations",  
Balasubramaniam, K. S. \& Simon, G. W., eds., 
ASP Conf. Ser., 68, 37 

\bibitem[Dainty \& Shaw(1974)]{dainty1974}
Dainty, J.~C. \& Shaw, R.  1974, Image Science, Academic (Press)

\bibitem[Denker et al.(1999)]{denker_etal1999}
Denker, C., Johannesson, A., Marquette, W., Goode, P.~R., Wang, H., \& Zirin, H. 1999, \solphys, 184, 87

\bibitem[Deslandres(1891)]{deslandres1891}
Deslandres, H. 1891, Comptes Rendus Acad. Sci. Paris, 131, 307

\bibitem[Deslandres \& D'Azambuja(1913)]{deslandres1913}
Deslandres, H. \& D'Azambuja, L. 1913, Comptes Rendus Acad. Sci. Paris, 157, 413

\bibitem[de Vaucouleurs(1968)]{devaucouleurs1968}
de Vaucouleurs, G. 1968, Applied Optics, 7, 1513


\bibitem[Ellerman(1919)]{ellerman1919}
Ellerman, F. 1968, \pasp, 31, 16

\bibitem[Ermolli et al.(1998)]{ermolli1998}
Ermolli, I., M. Fofi, Bernacchia, C., Berrilli, F., Caccin, B., Egidi, A., \& Florio, A. 1998, \solphys, 177, 1

\bibitem[Ermolli, Berrilli, \& Florio(2003)]{ermolli2003}
Ermolli, I., Berrilli, F., \& Florio, A. 2003, \aap, 412, 857

\bibitem[Ermolli et al.(2007)]{ermolli2007}
Ermolli,  I., Tlatov, A.~G., Solanki, S.~K., Krivova, N.~A., \& Singh, J. 2007, 
"The Physics of Chromospheric Plasmas",  Heinzel, P., Dorotovic, I. \& Rutten R.J., eds., 
ASP Conf. Series, 368, 533

\bibitem[Evershed(1911)]{evershed1911}
Evershed, W. 1911, \mnras, 71, 719

\bibitem[Fuller et al.(2005)]{fuller2005}
Fuller, N., Aboudarham, J., \& Bentley, R.D. 2005, \solphys, 227, 61

\bibitem[Gasperini, Mazzoni, \& Righini(2004)]{ga2004}
Gasperini, A., Mazzoni, M., \& Righini, A. 2004,
Giornale di Astronomia, 3, 23

\bibitem[Giorgi et al.(2005)]{giorgi2005}
Giorgi, F., Ermolli, I., Centrone, M., \& Marchei, E. 2005, Mem. SAIt., 76, 977

\bibitem[Godoli \& Righini(1950)]{gr1950}
Godoli, G. \& Righini, A. 1950, Mem. SAIt., 21, 4


\bibitem[Keller, Harvey \& The Solis Team(2003)]{kellerharvey2003}
Keller, C.U., Harvey, J.W., \& The Solis Team 2003, "Solar Polarization", Trujillo-Bueno J. \& Sanchez Almeida J., eds., 
ASP Conf. Series, 307, 13


\bibitem[Lefebvre et al.(2005)]{lefebvre2005} 
Lefebvre, S., Ulrich, R.~K., Webster, L.~S., Varadi, F., Javaraiah, J., Bertello, L., Werden, L., Boyden, J.~E., \& Gilman, P. 
2005, Mem. SAIt., 76, 862

\bibitem[Livingston \& Sheeley(2008)]{livingston2008}
Livingston, W. \& Sheeley, N.~R. 2008, \apj, 672, 1228

\bibitem[Makarov et al.(2004)]{makarov2004} 
Makarov, V.~I., Tlatov, A.~G., Singh, J., \& Gupta, S.~S. 2004, 
"Multi-Wavelength Investigations of Solar Activity", 
Stepanov, A. V., Benevolenskaya, E. E., \& Kosovichev, A. G., eds.,    
 IAU Symp. 223, 125

\bibitem[Marchei et al.(2006)]{marchei2006}
Marchei, E., Ermolli, I., Centrone, M., Giorgi, F., \& Perna, C. 2006, Mem. SAIt. S., 9, 51


\bibitem[Mickaelian et al.(2007)]{burkanian}
Mickaelian, A.~M., Nesci, R., Rossi, C., Weedman, D. et al. 2007, \aap, 464, 1177 

\bibitem[Mouradian \& Garcia(2007)]{mouradian2007} 
Mouradian, Z. \& Garcia, A. 2007, "The Physics of Chromospheric Plasmas", 368, 3, Heinzel, P., Dorotovic, I. \& Rutten R.J., eds., 
ASP Conf. Series, 368, 3


\bibitem[{Ortiz \& Rast(2005)}]{ortiz_rast2005}
Ortiz, A. \& Rast, M.~P. 2005, Mem. SAIt. 76, 4, 1018


\bibitem[Scherrer et al.(1995)]{scherrer1995}
Scherrer, P.~H., Bogart, R.~S., Bush, R.~I.  et al. 1995, \solphys, 162, 129

\bibitem[Schrijver et al.(1989)]{schrijver_etal1989}
Schrijver, C.~J., Ct\'e, J., Zwaan, C., \& Saar, S.~H. 1989, \apj, 337, 964

\bibitem[Skumanich, Smythe, \& Frazier(1975)]{skumanich1975}
Skumanich, A., Smythe, C., \& Frazier, E.~N. 1975, \apj, 200, 747


\bibitem[Ulrich et al.(2004)]{ulrich2004} 
Ulrich, R.~K., Webster, L.~S., Varadi, F., Javaraiah, J., Lefebvre, S., \& Gilman, P. 2004, AGU Fall Meeting Abstracts, A3


\bibitem[Walton et al.(1998)]{walton_etal1998}
Walton, S.~R., Chapman, G.~A., Cookson, A.~M., Dobias, J.~J., \& Preminger, D.~G. 1998, \solphys, 179, 31

\bibitem[Zharkova et al.(2003)]{zharkova_etal2003}
Zharkova, V.~V., Ipson, S.~S., Zharkov, S.~I., Benkhalil, A.~K., Aboudarham, J., \& Bentley, R.~D.
2003, \solphys, 214, 89

\end{thebibliography}
\end{document}